\documentclass[preprint,3p,twocolumn]{elsarticle}

\usepackage[colorlinks]{hyperref}
\usepackage{tabularx}
\usepackage{graphicx}
\usepackage{amssymb,amsmath,bm,tensor,braket}
\usepackage{xcolor}
\usepackage[varg]{txfonts}
\usepackage{enumerate}
\usepackage{mathtools}
\usepackage[utf8]{inputenc}
\usepackage[normalem]{ulem}
\usepackage{lineno,hyperref}
\usepackage{cuted}
\modulolinenumbers[5]

\journal{Physics Letters B}

\def\al{\alpha}
\def\be{\beta}
\def\ga{\gamma}

\def\th{\theta}

\def\ka{\kappa}
\def\la{\lambda}

\def\ph{\phi}

\def\ps{\psi}

\def\De{\Delta}

\def\Si{\Sigma}

\def\mn{{\mu\nu}}

\def\prt{\partial}
\def\pt{\phantom}

\def\sqr#1#2{{\vcenter{\vbox{\hrule height.#2pt
         \hbox{\vrule width.#2pt height#1pt \kern#1pt
         \vrule width.#2pt}
         \hrule height.#2pt}}}}

\newcommand{\beq}{\begin{equation}}
\newcommand{\eeq}{\end{equation}}
\newcommand{\bea}{\begin{eqnarray}}
\newcommand{\eea}{\end{eqnarray}}
\newcommand{\rf}[1]{(\ref{#1})}

\newcommand{\bM}{\begin{pmatrix}}
\newcommand{\eM}{\end{pmatrix}}

\begin{document}

\begin{frontmatter}

\title{The stealth Kerr solution in the bumblebee gravity}


\author[1]{Rui Xu}
\author[2]{Zhan-Feng Mai\corref{cor1}}\ead{zf1102@gxu.edu.cn}
\author[3]{Dicong Liang\corref{cor1}}\ead{dcliang@smu.edu.cn}

\address[1]{Department of Astronomy, Tsinghua University, Beijing 100084, China}
\address[2]{Guangxi Key Laboratory for Relativistic Astrophysics, School of Physical Science and Technology, Guangxi University, Nanning 530004,  China}
\address[3]{Department of Mathematics and Physics, School of Biomedical Engineering, Southern Medical University, Guangzhou, 510515, China}

\begin{abstract}
In this paper, we find Kerr solution accompanied with a nontrivial vector field as a solution to one of the simplest vector-tensor theories of gravity, namely the bumblebee model with an intriguing coupling constant between the Ricci curvature tensor and the vector field. We also demonstrate that the accompanied vector field can be generated via the Newman-Janis algorithm from a simple spherical vector field, which together with the Schwarzschild metric constitutes a solution to the same bumblebee model. It is probably the simplest example of a theory and its black-hole solutions for the Newman-Janis algorithm to hold except for general relativity.   
\end{abstract}


\end{frontmatter}

\section{Introduction}
\label{sec:intro}
The bumblebee gravity model extends general relativity (GR) with a vector field mimicking the kinetic behavior of the electromagnetic four-potential but having nonminimal couplings to the Ricci curvature tensor and the curvature scalar. 
The action reads
\begin{align}
S =&\, \frac{1}{2\ka} \int d^4x \sqrt{-g} \left(  R + \xi_1 B^\mu B^\nu R_\mn + \xi_2 B^\mu B_\mu R \right)  
\nonumber \\
& - \int d^4x \sqrt{-g} \left(\frac{1}{4} B^\mn B_\mn + V \right)  ,
\label{actionB}
\end{align}  
where $\ka=8\pi G$, $B_\mu$ is the vector field, $B_\mn := D_\mu B_\nu - D_\nu B_\mu$, and $D_\mu$ is the covariant derivative. 
The action differs from the Einstein-Maxwell theory due to the coupling terms controlled by the constants $\xi_1$ and $\xi_2$, and the potential term $V$ which is a function of the vector field in general.

A proper introduction to the original bumblebee gravity model, which focuses on its representative role as a gravity theory with possible spontaneous Lorentz symmetry breaking, is in Ref.~\cite{Kostelecky:2003fs}. 
Various properties of Lorentz symmetry breaking in the bumblebee model have been widely studied in \cite{Bluhm:2004ep,Bertolami:2005bh,Bailey:2006fd,Bluhm:2007bd,Bluhm:2008yt,Kostelecky:2010ze,Ovgun:2018xys,Liang:2022hxd,Bailey:2025oun, Zhu:2024qcm,Xu:2025heq,Ji:2024aeg,Shi:2025plr,AraujoFilho:2025hkm,AraujoFilho:2026yaj,AraujoFilho:2026oiy}. Futhermore, the study on the bumblebee black hole (BH) and other vacuum solution is also prosperous \cite{Casana:2017jkc,Xu:2022frb,Xu:2023xqh,Liang:2022gdk,Hu:2023vsg,Mai:2023ggs,Mai:2024lgk,Liu:2025oho,Li:2025itp,Li:2025tcd,AraujoFilho:2025rvn,Deng:2025uvp,Chen:2025ypx,Mustafa:2024mvx,Zhang:2023wwk,Liu:2022dcn,Filho:2022yrk,AraujoFilho:2024ykw,AraujoFilho:2024iox}. 
In this work, we will focus on an extraordinary analogy between a specific bumblebee model and GR with regard to BH solutions.  

The specific bumblebee model that we find interesting refers to the bumblebee model in Eq.~\rf{actionB} with $\xi_1=2\ka, \, \xi_2=0$ and $V=0$, because (i) a stealth Schwarzschild solution is found in this specific model in Ref.~\cite{Xu:2022frb}, 
and (ii) we now find it admit a stealth Kerr solution as well. 
In other words, the rotating BHs possess the Kerr metric, accompanied with a nontrivial vector field.
In addition, we will show that the vector field of the stealth Kerr solution can be derived from the vector field of the stealth Schwarzschild solution using the Newman-Janis algorithm.
Our work is firstly motivated by stimulating generalizations of the Newman-Janis algorithm and techniques in finding analytic solutions in other gravity theories. It is well known that most of BHs in our Universe is with rotation so that we expect our results can be the first step for testing bumblebee model through observational data, such as gravitational wave from LVK or black hole image from Event Horizon Telescope Collabration. 

\section{From the stealth Schwarzschild BH to Kerr solution}
\label{sec:pre}
Here, we directly present the stealth Kerr solution. The metric is most familiar,
\begin{align} 
ds^2 =&\, -\left( 1-\frac{2M\tilde r}{\Si} \right) d\tilde t^2 - \frac{4Ma\tilde r\sin^2\tilde\th}{\Si}d\tilde td\tilde r + \frac{\Si}{\De} d\tilde r^2 
\nonumber \\
& + \Si d\tilde\th^2 + \left[ \De + \frac{2M\tilde r\left(\tilde r^2+a^2\right)}{\Si} \right] \sin^2\tilde\th d\tilde\ph^2 ,
\label{kerr}
\end{align} 
where $\Si = \tilde r^2 + a^2 \cos^2\tilde\th$, $\De = \tilde r^2 + a^2 - 2M\tilde r$, and $M$ and $a$ are the mass and spin parameters. The vector field is nontrivial,
\begin{align}
b_\mu dx^\mu = \la_0 \left[ \left( 1-\frac{2M\tilde r}{\Si} \right) d\tilde t + \frac{2Ma\tilde r\sin^2\tilde \th}{\Si} d\tilde \ph \right],
\label{bumkerr}
\end{align}  
where we have denoted $B_\mu = b_\mu$, and $\la_0$ is a free constant.
It is straightforward to verify that the solution constituted by Eq.~\rf{kerr} and Eq.~\rf{bumkerr} satisfies the field equations of the umblebee model. It is obviously that the stealth Kerr solution share the same metric of Kerr black hole in GR. However, the stealth Kerr solution in bumblebee model is companied with a non-trivial vector field.  To clarify the specific model we are dealing with, we may as well review some basic equations and properties for the bumblebee BHs.

In this paper, we consider the case of $\xi_1=\xi, \, \xi_2=0$ and $V=0$. The equation of motion with respect to the specific action Eq.~\eqref{actionB} is 
\bea
G_\mn =  \ka \left( T_{b}\right)_\mn \quad D^\mu b_{\mn} + \frac{\xi}{\ka} b^\mu R_\mn  = 0,
\label{fieldeqs2}
\eea
where 
\begin{align}
\left( T_{b}\right)_\mn =&\, \frac{\xi}{2\ka} \Big[ g_\mn b^\al b^\be R_{\al\be} - 2 b_\mu b_\la R_\nu^{\pt\nu \la} - 2 b_\nu b_\la R_\mu^{\pt\mu \la}  
\nonumber \\
& - g_{\mn} D_\al D_\be ( b^\al b^\be ) - \Box_g ( b_\mu b_\nu )
\nonumber \\
& + D_\ka D_\mu \left( b^\ka b_\nu \right) + D_\ka D_\nu ( b_\mu b^\ka )   \Big]
\nonumber \\
& + b_{\mu\la} b_\nu^{\pt\nu \la} - \frac{1}{4} g_\mn  b^{\al\be} b_{\al\be}   .
\end{align}
The stealth Kerr solution will arise in the specific case of $\xi=2\ka$. Let us review the analytic stealth Schwarzschild black hole found in Ref.~\cite{Xu:2022frb}.
Denoting
\bea
ds^2 = -e^{2\nu}dt^2 + e^{2\mu} dr^2 + r^2 \left( d\th^2 + \sin^2\th \, d\ph^2 \right) ,
\label{ssmetric}
\eea
with $b_\mu = (b_t, b_r, 0, 0)$, one can find an analytic spherical solution to Eq.~\rf{fieldeqs2},
\begin{align}
\nu=& \, \frac{1}{2} \ln{\left(1-\frac{2M}{r} \right)} , 
\nonumber \\
\mu =& \, \mu_0 - \frac{1}{2} \ln{\left(1-\frac{2M}{r} \right)}, 
\nonumber \\
b_t =& \, \la_{0} + \frac{\la_{1}}{r}, 
\nonumber \\
b_r^2 =&\, e^{2\mu_{0}} \Bigg[ \frac{1}{\xi} \frac{\left(e^{2\mu_{0}}-1\right)r }{ r-2M} - \frac{\ka \la_1^2}{3\xi M (r-2M)}
\nonumber \\
& + \frac{\la_1^2 (2r-M) + 6\la_0\la_1 Mr + 6 \la_0^2 M^2 r}{3M (r-2M)^2} \Bigg] ,
\label{asol1}
\end{align} 
where $M$ is the mass parameter, and $\mu_0, \, \la_0, \, \la_1$ are three extra free parameters. 
The stealth Kerr solution at $a=0$ is the very special spherical solution,
\begin{align}
\nu =& \,-\mu = \frac{1}{2} \ln{\left(1-\frac{2M}{r} \right)} , 
\nonumber \\
b_t =& \, \la_{0} \left( 1 - \frac{2M}{r} \right), 
\nonumber \\
b_r =&\, 0 ,
\label{asol2}
\end{align} 
given by $\xi=2\ka, \, \mu_0 = 0$, and $\la_1 = -2M\la_0$ in Eq.~\rf{asol1}.  It shows that in the case of $\xi = 2\kappa$, there exist a spherical black hole, sharing the same metric with Schwarzschild BH in GR but with a nontrivial vector field. This originates from when considering the case of $\xi = 2\kappa$ and assuming that the metric is spherical or axially symmetric, the energy momentum tensor  $(T_{b})_{\mu\nu}$ is vanishing. This implies that $(T_{b})_{\mu\nu}$ contributed by the kinetic term of the vector field $\frac{1}{4} B^\mn B_\mn$, and by the non-minimal coupling term of the Ricci tensor and the vector field $ \xi_1 B^\mu B^\nu R_\mn$, cancel each other exactly. Thus, one finds that the Einstein tensor $G_{\mu\nu} = 0$, which leads to the emergence of the Schwarzschild and Kerr solutions. The stealth Kerr solution, Eq.~\eqref{kerr} is actually obtained through an ansatz, since it is derived by analogy from stealth Schwarzschild BH, and thus lacks mathematical rigor. We further find that this stealth Kerr solution can be constructed by the Newman-Janis algorithm.

\section{Newman-Janis algorithm and the stealth Kerr solution}
\label{sec:II}
In this section, we show that the stealth Kerr solution can be generated from the very special spherical solution given by $\xi=2\ka, \, \mu_0 = 0$ and $\la_1 = -2M\la_0$ via the Newman-Janis algorithm \cite{Newman:1965tw}. 
The metric part is of course already known in GR. The focus is on the vector field. We will construct the stealth Kerr solution along two paths: the original tetrad formalism \cite{Newman:1965tw} and the Giampieri's formalism \cite{Giampieri:1990}.
  
\subsection{Tetrad formalism}
The Newman-Janis algorithm in the tetrad formalism consists of the following three steps:
\begin{enumerate}
\item Introduce one of the Eddington–Finkelstein coordinates to express the null tetrad and the vector field. We use the outgoing Eddington–Finkelstein coordinate $u=t-r_*$, where $r_*$ is the tortoise coordinate defined via
\bea
\frac{dr_*}{dr} = \left( 1-\frac{2M}{r}\right)^{-1} , 
\label{tortcoord}
\eea  
so the null tetrad is
\bea
&& l^\mu \prt_\mu = \prt_r , 
\nonumber \\
&& n^\mu \prt_\mu = \prt_u - \frac{1}{2} \left( 1-\frac{2M}{r} \right) \prt_r, 
\nonumber \\
&& m^\mu \prt_\mu = \frac{1}{\sqrt{2}\, r} \left( \prt_\th + \frac{i}{\sin\th} \prt_\ph \right),
\nonumber \\
&& {\bar m}^\mu \bar \prt_\mu = \frac{1}{\sqrt{2}\, r} \left( \prt_\th - \frac{i}{\sin\th} \prt_\ph \right),
\eea
and the vector field of the stealth Schwarzschild BH in Eq.~\rf{asol2} surprisingly simplifies to
\bea
b^\mu \prt_\mu = -\la_0 \prt_u .
\label{bumss}
\eea

\item In $n^\mu$, take the replacement
\bea
\frac{1}{r} \rightarrow \frac{1}{2} \left( \frac{1}{r} + \frac{1}{\bar r} \right),
\label{rep1}
\eea
and assume $r$ to be complex.

\item Change the coordinates
\bea
&& u' = u-ia\cos\th,
\nonumber \\
&& r' = r+ia\cos\th,
\nonumber \\
&& \th'= \th,
\nonumber \\
&& \ph' = \ph ,
\eea
so 
\begin{align}
l^\mu \prt_\mu =&\, \prt_{r'},
\nonumber \\
n^\mu \prt_\mu =&\, \prt_{u'} - \frac{1}{2} \left[ 1 - M \left( \frac{1}{r} + \frac{1}{\bar r} \right) \right]\prt_{r'}, 
\nonumber \\
m^\mu \prt_\mu =&\, \frac{1}{\sqrt{2} \, r} \Big[ ia\sin\th \left(\prt_{u'}-\prt_{r'}\right) 
\nonumber \\
& + \prt_{\th'} + \frac{i}{\sin\th} \prt_{\ph'} \Big],
\nonumber \\
{\bar m}^\mu \bar \prt_\mu =&\, \frac{1}{\sqrt{2} \, \bar r} \Big[ -ia\sin\th \left(\prt_{u'}-\prt_{r'}\right) 
\nonumber \\
& + \prt_{\th'} - \frac{i}{\sin\th} \prt_{\ph'} \Big] ,
\label{nulltetradkerr1}
\end{align} 
and 
\bea
b^\mu \prt_\mu = -\la_0 \prt_{u'} .
\label{nullbumkerr1}
\eea
Note that the new coordinates $u'$ and $r'$ are assumed to be real.

\end{enumerate}

Now one can write down the metric corresponding to the null tetrad in Eq.~\rf{nulltetradkerr1}, 
\begin{align}
ds^2 =&\, -\left( 1-\frac{2Mr'}{\Si'}\right) du'^2 - 2du'dr' + \Si' d\th'^2
\nonumber \\
& + 2a\sin^2\th' dr'd\ph' -\frac{4M a r' \sin ^2\th'}{\Si'} du' d\ph'
\nonumber \\
&  + \left[\De' + \frac{2Mr'\left(r'^2+a^2\right)}{\Si'} \right] \sin^2\th' d\ph'^2 ,
\label{kerr2}
\end{align} 
and the associated vector field in Eq.~\rf{nullbumkerr1}
\begin{align}
b_\mu dx^\mu = &\, \la_0 \Bigg[ \left(1-\frac{2Mr'}{\Si'}\right) du' +  dr' 
\nonumber \\
& + \frac{2 Ma r' \sin ^2\th'}{\Si'} d\ph' \Bigg],
\label{bumkerr2}
\end{align}
where $\Si' = r'^2 + a^2\cos^2\th'$ and $\De'=r'^2+a^2-2Mr'$.
It is straightforward to verify that Eqs.~\rf{kerr2} and \rf{bumkerr2} constitute a solution to Eq.~\rf{fieldeqs2} for $\xi=2\ka$, and that a coordinate transformation 
\begin{align}
& \tilde t = u' + f_1(r'), 
\nonumber \\
& \tilde r = r' ,
\nonumber \\
& \tilde \th = \th' ,
\nonumber \\
& \tilde \ph = \ph' + f_2(r') ,
\end{align}
with 
\begin{align}
& \frac{d}{dr'} f_1(r') = \frac{r'^2+a^2}{\De'}, 
\nonumber \\
& \frac{d}{dr'} f_2(r') = \frac{a}{\De'} ,
\end{align}
recover the Kerr metric in Eq.~\rf{kerr} and the vector field in Eq.~\rf{bumkerr}. 

\subsection{Giampieri's formalism}
The Newman-Janis algorithm in Giampieri's formalism consists of the following four steps:
\begin{enumerate}
\item Introduce one of the Eddington-Finkelstein coordinates to express the line element and the vector field. We use $u=t-r_*$, where $r_*$ is the tortoise coordinate defined in Eq.~\rf{tortcoord}. The line element is
\begin{align}
ds^2 =&\, -\left( 1-\frac{2M}{r}\right) du^2 - 2 du dr 
\nonumber \\
& + r^2 \left( d\th^2 + \sin^2\th d\ph^2\right) ,
\label{metricss1}
\end{align}
and the vector field is shown in Eq.~\rf{bumss}, or equivalently
\bea
b_\mu dx^\mu = \la_0 \left[ \left( 1-\frac{2M}{r}\right) du + dr \right] .
\label{bumss1}
\eea

\item In Eqs.~\rf{metricss1} and \rf{bumss1}, take the replacement 
\begin{align}
\frac{1}{r} \rightarrow&\, \frac{1}{2} \left( \frac{1}{r} + \frac{1}{\bar r} \right),
\nonumber \\
r^2 \rightarrow &\, r \bar r,
\end{align}
and assume $r$ to be complex.

\item Change the coordinates
\begin{align}
u'= &\,u - ia\cos\ps,
\nonumber \\
r' = &\, r+ia\cos\ps,
\nonumber \\
\th' = &\, \th,
\nonumber \\
\ph' = &\, \ph ,
\end{align}
so 
\begin{align}
ds^2 =&\, -\left[ 1- M \left( \frac{1}{r} + \frac{1}{\bar r}\right) \right]du'^2 - 2 du' dr' 
\nonumber \\
& -2iMa\left( \frac{1}{r} + \frac{1}{\bar r}\right) \sin\ps du'd\ps 
\nonumber \\
& + 2ia\sin\ps dr' d\ps
\nonumber \\
& -a^2\left[ 1+ M \left( \frac{1}{r} + \frac{1}{\bar r}\right) \right]\sin^2\ps d\ps^2 
\nonumber \\
& + r \bar r \left( d\th^2 + \sin^2\th d\ph^2\right) ,
\label{metricss2}
\end{align}
and 
\begin{align}
b_\mu dx^\mu =&\, \la_0 \left[ 1- M \left( \frac{1}{r} + \frac{1}{\bar r}\right) \right] du' + \la_0 dr'
\nonumber \\
& + i\la_0 Ma \left( \frac{1}{r} + \frac{1}{\bar r}\right) \sin\ps d\ps.
\label{bumss2}
\end{align}

\item Slice the 5-dimensional spacetime using the ansatz
\bea
id\ps = \sin\ps d\ph',
\eea
and then replace $\ps \rightarrow \th'$. One gets the metric in Eq.~\rf{kerr2} and the vector field in Eq.~\rf{bumkerr2}.

\end{enumerate}

\section{No analytic rotating solutions found in the general case}
\label{sec:III}
It is unclear to us how to work out a rotating solution from the general spherical solution in Eq.~\rf{asol1}. 
The expression for $b_r$ is too complicated so that ambiguities arise when replacing the radial coordinate with certain combination of itself and its complex conjugate in the Newman-Janis algorithm. 
Noticing that the Newman-Janis algorithm succeeds for the special case of $\xi=2\ka, \, \mu_0=0$, and $\la_1 = -2M\la_0$, we can try to also apply it to the following relaxed cases:  
\begin{enumerate}
\item $\xi=2\ka, \, \mu_0=0$ so that 
\bea
b_r^2 = \left(\frac{\la_1+2M\la_0}{2M} \right)^2 \frac{2M}{r} \left(1-\frac{2M}{r}\right)^{-2} .
\eea    

\item $\xi=2\ka, \, \la_1=-2M\la_0$ so that 
\bea
b_r^2 = \frac{e^{2\mu_0}\left( e^{2\mu_0}-1\right)}{2\ka} \left(1-\frac{2M}{r}\right)^{-1} .
\eea    

\item $\mu_0=0, \, \la_1=-2M\la_0$ so that 
\bea
b_r^2 = \frac{\la_0^2\left(\xi-2\ka\right)}{3\xi} \frac{2M}{r} \left(1-\frac{2M}{r}\right)^{-1} .
\eea    

\end{enumerate}
One can verify that replacing $1/r$ using Eq.~\rf{rep1} for all the three relaxed cases does not produce rotating solutions to Eq.~\rf{fieldeqs2} unless they degenerate to the special case of $b_r=0$.

In fact, one can carry out the Newman-Janis algorithm using the general spherical solution in Eq.~\rf{asol1}. 
If we assume that the replacement in Eq.~\rf{rep1} still works for the metric and for the $b_t$ component, then the algorithm leads us to a definite metric and a vector field with only one undetermined function. We can follow Giampieri's formalism to show it:
\begin{enumerate}
\item Introduce the Eddington-Finkelstein coordinate $u=t-r_*$, where $r_*$ is defined via $dr_*/dr = e^{\mu-\nu}$. The line element is then
\begin{align}
ds^2 =&\, -e^{2\nu} du^2 - 2 e^{\mu+\nu} du dr 
\nonumber \\
& + r^2 \left( d\th^2 + \sin^2\th d\ph^2\right) ,
\label{gmetricss1}
\end{align}
and the vector field is
\bea
b_\mu dx^\mu = b_t du + \left( e^{\mu-\nu} b_t + b_r \right)dr.
\label{gbumss1}
\eea

\item In the metric part and $b_t$, take the replacement 
\begin{align}
\frac{1}{r} \rightarrow&\, \frac{1}{2} \left( \frac{1}{r} + \frac{1}{\bar r} \right),
\nonumber \\
r^2 \rightarrow &\, r \bar r.
\end{align}
In $b_r$, the corresponding replacement is unspecified, leading to an undetermined function at last.

\item Change the coordinates
\begin{align}
u'= &\,u - ia\cos\ps,
\nonumber \\
r' = &\, r+ia\cos\ps,
\nonumber \\
\th' = &\, \th,
\nonumber \\
\ph' = &\, \ph ,
\end{align}
so 
\begin{align}
ds^2 =&\, -e^{2\nu}du'^2 - 2e^{\mu+\nu} du' dr' 
\nonumber \\
& +2ia\left( e^{2\nu} - e^{\mu+\nu}\right) \sin\ps du'd\ps 
\nonumber \\
& + 2ia e^{\mu+\nu} \sin\ps dr' d\ps
\nonumber \\
& +a^2\left( e^{2\nu} - 2e^{\mu+\nu} \right) \sin^2\ps d\ps^2 
\nonumber \\
& + r \bar r \left( d\th^2 + \sin^2\th d\ph^2\right) ,
\label{gmetricss2}
\end{align}
and 
\begin{align}
b_\mu dx^\mu =&\, b_t \left(du'-ia\sin\ps d\ps\right) 
\nonumber \\
& + \left( e^{\mu-\nu} b_t + b_r \right)\left(dr'+ia\sin\ps d\ps\right).
\label{gbumss2}
\end{align}

\item Slice the 5-dimensional spacetime using the ansatz
\bea
id\ps = \sin\ps d\ph',
\eea
and then replace $\ps \rightarrow \th'$.

\end{enumerate}   
Noticing that $r = r'-ia\cos\ps \rightarrow r'-ia\cos\th'$, we have
\begin{align}
e^{2\nu} =&\, 1-\frac{2M}{r} \rightarrow 1-\frac{2Mr'}{\Si'},
\nonumber \\
b_t =&\, \la_0+\frac{\la_1}{r} \rightarrow \la_0+\frac{\la_1 r'}{\Si'} ,
\end{align}
where $\Si' = r'^2 + a^2\cos^2\th'$. Therefore, the resultant metric is 
\begin{align}
ds^2 =&\, -\left( 1-\frac{2Mr'}{\Si'} \right)du'^2 - 2e^{\mu_0} du' dr' 
\nonumber \\
& +2a\left( 1-\frac{2Mr'}{\Si'}- e^{\mu_0} \right) \sin^2\th' du'd\ph' 
\nonumber \\
& + 2 a e^{\mu_0} \sin^2\th' dr' d\ph'
\nonumber \\
& - a^2\left( 1-\frac{2Mr'}{\Si'} - 2e^{\mu_0} \right) \sin^4\th' d\ph'^2 
\nonumber \\
& + \Si' \left( d\th'^2 + \sin^2\th' d\ph'^2\right) ,
\label{gmetricss3}
\end{align}
which is Kerr when $\mu_0=0$. The resultant vector field can be written as
\begin{align}
b_\mu dx^\mu =&\, \left( \la_0+\frac{\la_1 r'}{\Si'} \right) \left(du'-a\sin^2\th' d\ph'\right) 
\nonumber \\
& + F_b(r', \th') \left(dr'+a\sin^2\th' d\ph'\right),
\label{gbumss3}
\end{align}
where $F_b(r', \th')$ is the undetermined function coming from $e^{\mu-\nu} b_t + b_r$.

Substituting the metric in Eq.~\rf{gmetricss3} and the vector field in Eq.~\rf{gbumss3} into the field equations in Eq.~\rf{fieldeqs2}, we find inconsistent equations for the undetermined function $F_b(r', \th')$, implying that the Newman-Janis algorithm fails for the general case. 
Note that the equations obtained for $F_b(r', \th')$ can be very complicated if one uses the Einstein field equations. 
A convenient way to see the inconsistency is using the vector field equation and set $\th'=\pi/2$ in the resultant equations.

\section{Conclusions}
\label{sec:disc}
We have demonstrated that a specific bumblebee model with $\xi_1=2\ka, \, \xi_2=0$ and $V=0$ admits a stealth Kerr solution, and that the stealth Kerr solution is generated exactly through the Newman-Janis algorithm from the stealth Schwarzschild solution in the same model. 
When the non-minimal coupling constant in the bumblebee model takes a general value rather than the specific value for the stealth bumblebee model, the Newman-Janis algorithm fails to generate any valid rotating solutions.  

The fact that such a specific bumblebee model has such an elegant BH solution makes it worth comparing with GR, or more strictly, with the vacuum Einstein-Maxwell theory which can also be regarded as the bumblebee model with $\xi=0$. In the full Einstein-Maxwell theory, the charge current is $J_Q^\mu = - D_\nu B^{\nu\mu}$ by virtue of the field equation. 
Now it can be used to define a charge
\begin{align}
Q : =&\, -\frac{ 1 }{4\pi} \sqrt{\frac{\ka}{2}} \int_{\Si} d^3x \sqrt{\ga} \,n_\mu J_Q^\mu , 
\end{align}
 in the bumblebee model, and there is also the conservation equation 
\bea
D_\mu J_Q^\mu = D_\mu D_\nu B^\mn = 0.
\eea
Using the stealth Kerr solution, one finds the charge being $Q=-\sqrt{2\ka}\, \la_0 M$. 
In this sense, the stealth Kerr solution is a charged rotating BH solution. 
In the Einstein-Maxwell theory, the charged rotating BH solution is the Kerr-Newman metric \cite{Newman:1965my}, which is more involved compared with the Kerr metric. 
It is a surprise that the nonminimal coupling in this specific bumblebee model actually simplifies the BH spacetime. Table~\ref{compbhsols} shows the comparison.   
Let us mention that this specific bumblebee model not only has such an elegant BH solution, but also leads to a very simple FLRW solution in cosmology \cite{Xu:2025heq}. 

It is a pity that the efforts for bumblebee gravity to replace the standard $\Lambda$CDM model failed \cite{Xu:2025heq}. 
But the idea that certain nonminimal coupling between the metric tensor and a vector field can cancel the effect of the vector field itself and make the vector field stealth in a strong-gravity regime, is interesting and appealing in constructing theories that unify gravity and electromagnetism. 

\begin{table}
\caption{Charged black holes in GR and in the stealth bumblebee model. }
\begin{tabular}{m{1.6cm}m{2.4cm}m{2.5cm}}
\hline\hline
 Theory & Charged BH & Charged rotating BH \\
\hline
GR & Reissner-Nordstr\"om  & Kerr-Newman   \\
\hline
Stealth bumblebee & Schwarzschild  & Kerr   \\
\hline
\end{tabular}
\label{compbhsols}
\end{table}

The stealth Schwarzschild BH solution was first discovered by Babichev \&  Charmousis \cite{Babichev:2013cya}, then subsequently identified in various scalar-tensor theories \cite{Kobayashi:2014eva,Babichev:2017guv,Motohashi:2018wdq,BenAchour:2018dap,Minamitsuji:2018vuw,Bernardo:2019yxp,Takahashi:2020hso}. This intriguing solution has been found in vector–tensor theories \cite{Heisenberg:2017xda,Heisenberg:2017hwb,Chagoya:2017ojn,Minamitsuji:2021gcq}, and scalar-vector-tensor theories \cite{Skordis:2024wlo} as well.
By extending the spherical solution to axisymmetric spacetimes, the stealth Kerr solution has also been discovered in modified gravity theories \cite{Cisterna:2016nwq,Charmousis:2019vnf}.  
Unlike the exact GR solutions, the thermodynamics of stealth solutions exhibit distinct behaviors \cite{Chagoya:2023ddb,Bakopoulos:2024zke}. 
Moreover, perturbations around the same background evolve differently in GR and in modified gravity theories \cite{Ogawa:2015pea,Kase:2018voo,deRham:2019gha,Khoury:2020aya,Bernardo:2020ehy,Takahashi:2021bml,Kobayashi:2025evr}. 
From an observational perspective, these differences could potentially allow us to distinguish modified gravity theories from GR through their quasi normal modes \cite{Mukohyama:2023xyf,Sirera:2024ghv,Charmousis:2025xug}.
It would be valuable to investigate the thermodynamics, perturbation behaviors and other properties of the stealth Kerr BH in the bumblebee model, and we leave these studies to the future work.

\section*{Acknowledgements}
Z.~F.~Mai is supported by National Key R\&D Program of China (2024YFA1611700). This work was also supported by the National Natural Science Foundation of China (Grants No.~12405070, 12405065, 12465013) and the Guangxi Talent Program (``Highland of Innovation Talents'').

\bibliography{mybibfile}

\end{document}